\documentclass[english,12pt]{article}
\usepackage{amsmath,amsfonts,amssymb,amsbsy,amstext}
\usepackage[top=3cm, bottom=2cm, left=2cm, right=2cm]{geometry}
\usepackage{graphicx}
\usepackage{color}
\usepackage[english]{babel}
\usepackage{subfig}
\usepackage[numbers,compress]{natbib}
\usepackage{tablefootnote}

\newcommand{\beq}{\begin{eqnarray}}
\newcommand{\eeq}{\end{eqnarray}}
\newcommand{\non}{\nonumber\\}

\newcommand{\p}{\partial}

\newtheorem{conjecture}{Conjecture}

\begin{document}

\begin{titlepage}
\def\thefootnote{\fnsymbol{footnote}}
\phantom{.}\vspace{-2cm}
\bigskip
\bigskip
\bigskip

\begin{center}
{\Large {\bf Some exact Bradlow vortex solutions}}

\bigskip
{\large Sven Bjarke Gudnason${}^1$\footnote{\texttt{bjarke(at)impcas.ac.cn}} 
and Muneto Nitta${}^2$\footnote{\texttt{nitta(at)phys-h.keio.ac.jp}}}
\end{center}

\renewcommand{\thefootnote}{\arabic{footnote}}

\begin{center}
\vspace{0em}
{\em {${}^1$Institute of Modern Physics, Chinese Academy of Sciences,
  Lanzhou 730000, China\\
${}^2$Department of Physics, and Research and Education Center for Natural
Sciences, Keio University, Hiyoshi 4-1-1, Yokohama, Kanagawa 223-8521,
Japan 

\vskip .4cm}}

\end{center}

\vspace{1.1cm}

\noindent
\begin{center} {\bf Abstract} \end{center}

We consider the Bradlow equation for vortices which was recently found
by Manton and find a two-parameter class of analytic solutions in
closed form on nontrivial geometries with non-constant curvature.
The general solution to our class of metrics is given by a
hypergeometric function and the area of the vortex domain by the
Gaussian hypergeometric function.

\vfill

\begin{flushleft}
{\today}
\end{flushleft}
\end{titlepage}

\section{Introduction}

Vortices are codimension-two solitons and are most famous for their
existence in type II superconductors in the presence of an external
magnetic field, where they carry quantized fluxes through the
superconducting material.
The effective theory for the latter effect is the Ginzburg-Landau
theory or -- from the point of view of the soliton -- equivalently the
Abelian Higgs model.
The type II vortices are described by two partial differential
equations (PDEs) -- one for the scalar field (order parameter) and
one for the gauge field -- as well as a parameter
$\beta=m_h/m_\gamma>1$, which is the ratio of the scalar mass to the
photon mass. 
Critically coupled vortices, also called BPS vortices, have $\beta=1$ 
and their two PDEs can be reduced to a single PDE which is known as
the Taubes equation \cite{Taubes:1979tm}
\beq
-\nabla^2 u = 1 - e^{2u} - ({\rm delta\ functions}),
\nonumber
\eeq
where $u=\log|\phi|$ is the logarithm of the modulus of the order
parameter. 
Although the existence and uniqueness have been proven for the Taubes
equation \cite{Taubes:1979tm}, no analytic solutions are known
on the flat plane, $\mathbb{R}^2$.
On the hyperbolic plane, $\mathbb{H}^2$, however, Witten found exact
vortex solutions \cite{Witten:1976ck}\footnote{Witten considered
  instanton solutions by splitting $\mathbb{R}^4$ into
  $\mathbb{H}^2\times S^2$ and constructed a nontrivial flux on the
  hyperbolic plane. } by adjusting the constant negative curvature of
the hyperbolic plane so as to effectively cancel the constant (vacuum
expectation value of the scalar field) in the Taubes equation. 
This way both the hyperbolic plane and the vortex scalar field are
described by Liouville's equation.
Only a few other analytic vortex solutions are known (see
e.g.~Ref.~\cite{Manton:2009ja}) and one property they have in common
-- to the best of our knowledge -- is that they exist on manifolds of 
constant Gaussian curvature. 

Other vortex equations of similar type to Taubes equation are the
Jackiw-Pi equation \cite{Jackiw:1990tz,Jackiw:1990mb} and the Popov
equation \cite{Popov:2012av}; both of which possess exact analytic
solutions and again on manifolds of constant Gaussian curvature.
The Jackiw-Pi equation comes about naturally in some non-relativistic
Chern-Simons theories \cite{Jackiw:1990tz,Jackiw:1990mb,Dunne:1990qe},
whereas the Popov equation describes vortices on a 2-sphere that are
realized as Yang-Mills instantons on a manifold
$S^2\times\mathbb{H}^2$ (the vortices here are situated on the
2-sphere, whereas in Witten's solution they are situated on the
hyperbolic plane). 
In a recent paper, Manton considered extending the vortex equations to
nine different types -- five of which can possess vortices -- and
thereby found two new ones \cite{Manton:2016waw}; one of them was
dubbed the Bradlow equation and the other remained an unnamed
equation.
The unnamed equation is, however, not really new as it was found by
Ambj\o rn and Olesen in
Ref.~\cite{Ambjorn:1988fx,Ambjorn:1988tm}\footnote{We thank 
  P.~Olesen for pointing this out. } and it describes
$W$-condensation giving rise to a periodic vortex lattice in the
electroweak theory. 
We will thus call the equation the Ambj\o rn-Olesen-Manton equation. 
The fact that it resembles the normal vortex equation but with
opposite sign of the last term can be interpreted as the effect of
anti-screening of the $W$-boson as opposed to the normal screening
effect in Abelian theories \cite{Ambjorn:1988fx,Ambjorn:1988tm}.
In the flat plane, $\mathbb{R}^2$, the Ambj\o rn-Olesen-Manton
equation does not have a single vortex solution, but instead a lattice
of vortices.
This can also be seen from the fact that the equation does not
possess a (constant) vacuum state, i.e.~a value of the field where the
Laplacian vanishes.
The equation does however have fixed points which are described by
analytic solutions presented in Ref.~\cite{Manton:2016waw}.
Finally, the Ambj\o rn-Olesen-Manton equation has also been suggested
recently to play a role in a non-Abelian vector bootstrap mechanism
generating a primordial magnetic field \cite{Olesen:2017mcu}. 

The other new vortex equation -- the Bradlow equation -- is remarkably
simple as it simply equates the magnetic flux with a constant. The
vortex field is thus energetically absent, but its zeros still
specify the positions of the vortices inside this constant magnetic
field.
The name of the Bradlow equation was coined in
Ref.~\cite{Manton:2016waw} due to the similarity of the equation with
the Bradlow bound \cite{Bradlow:1990ir}, which however was formulated
for the Taubes equation and limits the number of vortices that can
exist on a compact manifold. 
Although applications for the Bradlow equation may not seem immediate, 
we would like to consider this as an approximation to a physical
system. 
Bose-Einstein condensates with constant magnetic fields exist
experimentally \cite{Lin:2009} and vortices -- global vortices
(i.e.~ungauged vortices) -- are created in such a way that the
magnetic field is constant even around the vortices, and so perhaps
the Bradlow equation can be used as a rough description in this case. 

In this paper, we consider the Bradlow equation and construct
analytic solutions in closed form for a two-parameter family of
metrics with non-constant Gaussian curvature; these are the first
analytic solutions on manifolds with a non-constant curvature.
The solutions are rather simple due to the fact that the Bradlow
equation is also rather simple and they consist of the vortex
positions, an overall normalization factor and a nontrivial function
whose solution is given in terms of a hypergeometric function.
We also tie together a physical aspect of the system, namely matching
the magnetic flux with the vortex number. In the case of the Taubes
equation, this is automatic, but for the Bradlow equation, a
restriction on the domain (area) is necessary for the double integral
of the equation to hold. 

The paper is organized as follows. In Sec.~\ref{sec:Bradloweq} we
introduce the Bradlow equation in perspective to the other
possibilities of the same type of equation.
In Sec.~\ref{sec:toymodel} we introduce a toy model giving rise to the 
Bradlow equation and discuss the boundary conditions.
Then in Sec.~\ref{sec:sols} we first solve the Bradlow equation on a
flat disc and then present the main result of the paper, namely the
solutions on curved manifolds with non-constant curvature. 
Sec.~\ref{sec:discussion} concludes with a discussion and outlook. 
Finally, in the appendix, we contemplate the uniqueness in the cases
of the modified vortex equations, useful also for the Bradlow
equation.

\section{The Bradlow equation}\label{sec:Bradloweq}

We start by introducing the class of vortex equations to which the
Bradlow equation belongs and from which it was discovered. 
They are given by \cite{Manton:2016waw},
\beq
-\frac{1}{\Omega_0}\nabla^2u = - C_0 + C e^{2u} 
- \frac{2\pi}{\Omega_0}\sum_{i=1}^N \delta^{(2)}(z - z_i),
\label{eq:vtx}
\eeq
where the constants $C_0$ and $C$ are given in
Table \ref{tab:vtxconstants}, $\{z_i\}$, $i=1,\ldots,N$, are vortex
positions, and $N$ is the total number of vortices.
The manifold on which the vortices and the above equations are defined 
is denoted by $M_0$ and has the metric factor $\Omega_0(z,\bar{z})$,
where the metric is defined by
\beq
ds^2 = dt^2 - \Omega_0 \left[(dx^1)^2 + (dx^2)^2\right]
= dt^2 - \Omega_0 dz d\bar{z},
\eeq
the complex coordinate is defined as $z=x^1+ix^2$,
$\Omega_0^{-1}\nabla^2$ is the covariant Laplacian on $M_0$ and
$\nabla^2=\p_{x^1}^2+\p_{x^2}^2=4\p_{\bar{z}}\p_z$. 
As explained in Ref.~\cite{Manton:2016waw}, an overall factor of
$\{C_0,C\}$ can be scaled away in Eq.~\eqref{eq:vtx} by rescaling the
metric factor $\Omega_0$ while $C$ can independently be scaled by a
shift in $u$ and thus nine distinct equations are given by $C_0$, $C$
taking a value in $\{-1,0,1\}$.
Of the nine possibilities, four do not allow for a vortex with zeroes
and a positive magnetic flux. 

\begin{table}[!htp]
  \caption{Vortex equation constants $C_0$ and $C$ for five different
    theories.}
  \label{tab:vtxconstants}
  \begin{center}
  \begin{tabular}{rr||lc}
    $C_0$ & $C$ & name & analytic solutions on\\ 
    \hline\hline
    $-1$ & $-1$ & Taubes & $\mathbb{H}^2$\\
    $0$  & $1$  & Jackiw-Pi & $\mathbb{R}^2$, $T^2$\\
    $1$  & $1$  & Popov & $S^2$ \\
    $-1$ & $0$  & Bradlow & $\mathbb{H}^2$\\
    $-1$ & $1$  & Ambj\o rn-Olesen-Manton & $\mathbb{H}^2$
  \end{tabular}
  \end{center}
\end{table}

We will now review the Bradlow integral relation for the generic case 
\cite{Manton:2016waw} by integrating Eq.~\eqref{eq:vtx} to
\beq
2\pi N = -C_0 A + C\int_{M_0} d^2x\; \Omega_0 e^{2u},
\label{eq:Bradlow_bound}
\eeq
where $N$ is the vortex number and $A$ is the area of $M_0$.
The left-hand side of Eq.~\eqref{eq:Bradlow_bound} is the topological
number, which we will take to be finite here; therefore two situations
arise: If the area $A$ is infinite, this puts a constraint on the
field $u$ to reach the vacuum
\beq
u \to u_{\infty} = \frac{1}{2}\log\frac{C_0}{C},
\label{eq:BC}
\eeq
at a specific rate (as does the equation of motion of course); this
equation is not valid for the Ambj\o rn-Olesen-Manton equation, but it
is for all the other equations. 
In the cases where either $C_0$ or $C$ vanishes, the above vacuum
expectation value should be taken as the limit of the latter
constant being sent to zero. 

If, on the other hand, the area $A$ is finite ($M_0$ is compact),
then three cases arise:
\begin{itemize}
\item $C_0=-1$: this is the normal Bradlow bound for the Taubes case:
  we get
  \beq
  N \leq \frac{A}{2\pi}, \qquad
  N = \frac{A}{2\pi}, \qquad
  N \geq \frac{A}{2\pi},
  \eeq
  for the Taubes, the Bradlow and the Ambj\o rn-Olesen-Manton
  equation, respectively, as $C$ is $-1,0,1$, respectively. 
\item $C_0=0$: in this case, the equation directly relates the vortex 
  number and the integral of the scalar field (squared)
  \beq
  N = \frac{1}{2\pi}\int_{M_0} d^2x\; \Omega_0 e^{2u}.
  \eeq
  Note: in this case, $C$ cannot vanish and since we take the vortex
  number to be positive, $C=1$ is the only possibility: i.e.~the
  Jackiw-Pi equation.
\item $C_0=1$: in this case the area cannot be too large and $C=1$ is
  required to get a positive vorticity; we can write an upper bound
  for the area
  \beq
  A < \int_{M_0} d^2x\;\Omega_0 e^{2u}.
  \eeq
  This upper bound for the area is valid for the Popov equation. 
\end{itemize}

In the remainder of the paper we will concentrate on the Bradlow
equation
\beq
-\nabla^2 u = \Omega_0
- 2\pi\sum_{i=1}^N \delta^{(2)}(z - z_i),
\label{eq:Bradlow}
\eeq
for which the topological vortex number, $N$, is related to the area
of the manifold or integration domain as
\beq
N = \frac{A}{2\pi}.
\label{eq:NArel}
\eeq
In the next section we will illustrate a toy model and set up the
boundary conditions.

\section{Toy model and boundary conditions}\label{sec:toymodel}

It will prove instructive to define a prototype theory giving rise to
the Bradlow equation.
Let us define the following static energy
\begin{align}
  E &= \int_{M_0} d^2x\; \Omega_0\left\{
  \frac{1}{2e^2\Omega_0^2} F_{12}^2
  + \Omega_0^{-1} |D_a\phi|^2
  + \Omega_0^{-1} |\phi|^2 F_{12}
  + \frac{1}{2}e^2 v^4 \right\}
  \non
&= \int_{M_0} d^2x\; \Omega_0\left\{
  \frac{1}{2e^2}\left(\Omega_0^{-1}F_{12} + e^2 v^2\right)^2
  + \Omega_0^{-1}|D_1\phi + i D_2\phi|^2
  - i\Omega_0^{-1} \epsilon^{a b} \p_a(\bar{\phi}D_b \phi)
  \right\} \non
&\phantom{=\ }
  - v^2 \int_{M_0} d^2x\; F_{12},
  \label{eq:toymodel}
\end{align}
where $F_{12}=\p_1A_2-\p_2A_1=B$ is the magnetic field in the plane,
$D_a=\p_a+iA_a$, is the (gauge) covariant derivative, $A_a$ is the
Abelian (U(1)) gauge field, the indices $a,b$ run over $1,2$; $\phi$
is a complex scalar field and $e>0$, $v>0$ are constants,
respectively, gauge coupling constant and vacuum energy (cosmological
constant). 

Let us emphasize that this theory is just a toy model and we take the
Bradlow equation as the defining equation. Therefore any other theory
giving rise to the Bradlow equation can be considered equally valid. 

Working with the above stated theory, we can read off the Bogomol'nyi
equations as
\begin{align}
  D_1\phi + i D_2\phi &\equiv 2D_{\bar{z}}\phi = 0,
  \label{eq:BPS1}\\
  -\frac{1}{\Omega_0} F_{12} &= m^2,
  \label{eq:BPS2}
\end{align}
where we have defined $m\equiv e v>0$.
In addition to the above equations, if we impose that $D_a\phi\to 0$
at the boundary of $M_0$ such that the total derivative in the energy
vanishes, then the total energy is
\beq
E = v^2 m^2 A = e^2 v^4 A,
\eeq
where $A$ is the total area of $M_0$: if $M_0$ is compact, then the
total energy is finite, if not then it is infinite.

Let us start by solving the first BPS equation,
i.e.~Eq.~\eqref{eq:BPS1},
\beq
D_{\bar{z}}\phi = \p_{\bar{z}}\phi + i A_{\bar{z}}\phi 
= \big(-\p_{\bar{z}}\log s + i A_{\bar{z}}\big) s^{-1} \phi_0,
\qquad\Rightarrow
A_{\bar{z}} = -i\p_{\bar{z}}\log s,
\eeq
where we have defined 
\beq
\phi(z,\bar{z}) \equiv s^{-1}(z,\bar{z}) \phi_0(z),
\label{eq:phi0def}
\eeq
where $s(z,\bar{z})$ is everywhere regular and $\phi_0(z)$ is
holomorphic and contains all zeros of the field $\phi$.
Calculating now the field strength, we get
\beq
F_{12} = 2i F_{\bar{z}z} = - 2\p_{\bar{z}}\p_z \log |s|^2
= 2\p_{\bar{z}}\p_z\log |\phi|^2
- 2\pi \sum_{i=1}^N \delta^{(2)}(z - z_i),
\eeq
which by insertion into the other BPS equation \eqref{eq:BPS2} yields
the Bradlow equation
\beq
-\frac{1}{\Omega_0} \nabla^2 u = 1
-\frac{2\pi}{\Omega_0}\sum_{i=1}^N\delta^{(2)}(z - z_i),
\label{eq:Bradlow_again}
\eeq
where $\nabla^2=4\p_{\bar{z}}\p_z$, we have defined
$u \equiv \frac{1}{2}\log |\phi|^2$, and rescaled the mass squared,
$m^2$, into the conformal factor of the metric,
$\Omega_0\to\Omega_0/m^2$. 

Let us consider carefully what assumptions go into the statement that
the vortex number is proportional to the area of $M_0$, see
Eq.~\eqref{eq:NArel}.
Inserting Eq.~\eqref{eq:phi0def} into Eq.~\eqref{eq:Bradlow_again}, we
get 
\beq
\frac{1}{2\Omega_0} \nabla^2 \log |s|^2 = 1,
\eeq
where $s(z,\bar{z})$ is everywhere regular.
If we now integrate the above equation over $M_0$, we get
\begin{align}
A
&= \frac{1}{2}\int_{M_0} d^2x\;  \nabla^2 \log |s|^2
= i\int_{M_0} dz\wedge d\bar{z}\; \p_{\bar{z}}\p_z\log |s|^2
= -i\oint_{\p M_0} dz\; \log |s|^2 \non
&= -i\oint_{\p M_0} dz\; \log |z|^{2k}
= 2\pi k, \label{eq:Arelk}
\end{align}
where we have used Green's theorem and in the second line we have
assumed the boundary condition 
\beq
\lim_{z\to z_{\p M_0}} |s|^2 = |z|^{2k},
\eeq
where $z_{\p M_0}$ is the coordinate lying on the boundary of $M_0$. 
Hence we see that the area of the manifold $M_0$ is related to the
winding of the gauge field $A_a$.

Let us first discuss the case of $M_0$ being noncompact; in
particular take $M_0=\mathbb{R}^2$.
We will assume that the vortex under study is topological, which means
that
\beq
\lim_{|z|\to\infty}|\phi| = {\rm const} >0,
\eeq
and hence we get that $|s^{-1}\phi_0|\sim |z|^{N-k}$ forces
$N=k$.\footnote{If $N<k$, $\lim_{|z|\to\infty}|\phi|=0$ and this is
  called a non-topological vortex.}
We will then consider the contribution to the energy
\eqref{eq:toymodel} from the total derivative term.
In this case ($M_0=\mathbb{R}^2$) we need to impose the condition that
$D_a\phi$ goes to zero at the boundary of $M_0$ so that the total
derivative term in the energy \eqref{eq:toymodel} vanishes.
Since $D_{\bar{z}}\phi=0$ everywhere, this means that we need to
impose the condition
\beq
0 = \lim_{|z|\to\infty} D_z\phi
= \left(-\p_z\log |s|^2 + \p_z\log\phi_0\right) s^{-1} \phi_0
\sim \left(-\frac{k}{z} + \frac{N}{z}\right) s^{-1} \phi_0.
\eeq
Note that from Eq.~\eqref{eq:Arelk}, we get that for an infinite area,
$k$ is also infinite and therefore, for the above covariant derivative
to vanish at the boundary of $M_0$, we need $N=k$. 
Thus, we again get $k=N$ and hence we come to the conclusion that
\beq
2\pi N = A,
\eeq
for $M_0$ being the flat infinite plane, $\mathbb{R}^2$.
Note that $N$ is required to be infinite due to the infinite area. 
The fact that the gauge field (formally) winds $N$ times and the
scalar field has exactly $N$ zeros (counted with multiplicity,
i.e.~they may be coincident) is in fact very natural for vortices.
Nevertheless, depending on the underlying theory, this is not strictly
necessary in the case of the Bradlow vortices, whereas in the case of
the Taubes equation, the vacuum of the scalar field also forces
$|s|=|\phi_0|$ at the boundary and hence $k=N$.
Therefore in the case of the Bradlow vortex, one could in principle
contemplate the situation where $k\neq N$, but that would imply that
the covariant derivative of the scalar field cannot go to zero at the
boundary of $M_0$ (for $N>k$, and the scalar field will diverge at
infinity) or that the vortices are not topological (for $N<k$). 
In particular for $M_0=\mathbb{R}^2$ and $N>k$, that would imply two
separate diverging contributions to the energy in the toy model; one
from the magnetic field and one from the total derivative term. 

Let us now discuss a compact case, in particular the flat disc,
$M_0=\mathbb{D}^2$ with radius $R$.
If we still demand that $D_z\phi=0$ for $|z|=R$, then we get
\beq
\left.-\p_z\log |s|^2 + \p_z\log \phi_0(z)\right|_{|z|=R}
= \left.-\p_z\log |s|^2 + \p_z\log |\phi_0(z)|^2\right|_{|z|=R}
= \left.\p_z u\right|_{|z|=R} = 0,
\eeq
where in the second equality we have used the holomorphicity of
$\phi_0(z)$ to add $\log\overline{\phi_0(z)}$ in the derivative.
This condition, however, cannot in general be satisfied by solutions
to the Bradlow equation.
Nevertheless, in the case of compact $M_0$, the boundary term in
the energy \eqref{eq:toymodel} does not give a diverging contribution
due to the finite size of $M_0$ and thus finite circumference.
Therefore one boundary condition we can choose to impose on the disc
is $u(R)=0$.
However, other boundary conditions may be equally reasonable,
depending on the desired behavior of the solution on the boundary
$\p M_0$.

In view of the above discussion, we will restrict the rest of the
paper to cases where $M_0$ is a compact manifold.

\section{Bradlow vortex solutions}\label{sec:sols}

\subsection{Bradlow vortices on a flat disc}

In this section, we will gain some intuition by studying the Bradlow
equation on the flat disc. 
On a disc with vanishing curvature $(\Omega_0=1$) and radius $R$, the
Bradlow vortex is the solution to the Bradlow equation
\eqref{eq:Bradlow} and it has the analytic solution with axial
symmetry 
\beq
u = -\frac{r^2}{4} + u_0 + \frac{N}{2} \log r^2,
\label{eq:Bradlow_radial}
\eeq
where $r=|z|$ is the radial coordinate, $u_0$ is a constant and $N$
is the vortex number.
Let us impose the boundary condition that $u(R)=0$, yielding
\beq
u = \frac{R^2-r^2}{4} + \frac{N}{2} \log\frac{r^2}{R^2},
\label{eq:Bradlow_radial_BC}
\eeq
where we have adjusted $u_0$ to make $u$ match its boundary
conditions.

Allowing for the vortices to have generic positions, the most general
solution is
\beq
u = -\frac{|z|^2}{4} + u_0 + \frac{1}{2}\sum_{i=1}^N\log|z-z_i|^2
+ g(z) + \overline{g(z)},
\label{eq:flatsol}
\eeq
which reduces to Eq.~\eqref{eq:Bradlow_radial} when all $z_i=0$ and
$g$ are set to zero. 
Imposing the boundary condition $u(R)=0$ is however highly nontrivial
in the this general case.
The constant $u_0$ can still be fixed to $R^2/4$ so that $u(R)=0$
except for the contribution due to the logarithms.
Naively to cancel the contribution coming from the logarithms, one
would guess that subtracting
\beq
\frac{1}{2}\sum_{i=1}^N\log|R e^{i\theta}-z_i|^2 =
\frac{1}{2}\sum_{i=1}^N\log\left|\frac{R z}{|z|}-z_i\right|^2,
\eeq
from the solution would make it satisfy the boundary condition
$u(R)=0$. However, the above term does not vanish when acted upon by
the Laplacian due to the non-holomorphic $1/|z|$ necessary for giving
the phase factor $e^{i\theta}$.
As we can see from the general solution \eqref{eq:flatsol}, we can
adjust $u_0$ and $g(z)$ to make the solution satisfy the boundary
condition $u(R)=0$. However, a holomorphic function which cancels the
contribution due to the logarithms at $|z|=R$ will also cancel the
logarithmic singularity defining the vortex center.
Hence we make the following conjecture.
\begin{conjecture}
The only solution satisfying the Bradlow equation \eqref{eq:Bradlow}
on the flat disc, $\mathbb{D}^2$, with a finite radius $R<\infty$ and
the boundary condition $u(R)=0$, is the axially symmetric solution
\eqref{eq:Bradlow_radial_BC} where all $z_i=0$, $\forall i$.
\label{conjecture1}
\end{conjecture}

Of course other boundary conditions than $u(R)=0$ may be chosen.
An alternative, is to choose the boundary conditions such that the
contribution due to the metric in the Bradlow equation vanishes at the 
boundary of the disc, but neglecting the nontrivial function due to
the logarithms. From this point of view, one can choose the positions
of the vortices, but the solution on the boundary does not satisfy any
simple boundary conditions. 
We can write such a solution as
\beq
u = \frac{R^2 - |z|^2}{4} + \frac{1}{2}\sum_{i=1}^N\log|z-z_i|^2.
\label{eq:flatsol_u0fixed}
\eeq

One can come arbitrarily close to the boundary condition $u(R)=0$ if
the size of the disc is parametrically larger than the distance from
the vortices to the center of the disc, in particular, $R\gg|z_i|$,
$\forall i$. In this case, we can fix $u_0$ as follows 
\beq
u = \frac{R^2 - |z|^2}{4}
+ \frac{1}{2}\sum_{i=1}^N\log\frac{|z-z_i|^2}{R^2}.
\label{eq:flatsol_u0fixed2}
\eeq
In this case at the boundary, we have
\beq
u(|z|=R) =
\frac{1}{2}\sum_{i=1}^N
\log\left|\frac{z}{|z|} - \frac{z_i}{R}\right|^2
\simeq -\frac{1}{2}\sum_{i=1}^N
\left[\frac{z_i|z|}{R z} + \frac{\bar{z}_i|z|}{R\bar{z}}
+ \mathcal{O}\left(\frac{|z_i|^2}{R^2}\right)\right],
\eeq
and hence for $R$ parametrically bigger than $|z_i|$ the discrepancy
between the above solution with the boundary condition $u(R)$ can
become arbitrarily small.
Nevertheless, for any finite size disc, the only solution strictly
satisfying $u(R)=0$ is conjectured to be
Eq.~\eqref{eq:Bradlow_radial_BC}, see Conjecture \ref{conjecture1}.

Since the vortex number is related to the area of the disc by the
Bradlow equation \eqref{eq:Bradlow}, the radius of the disc is
determined as
\beq
N = -\frac{1}{2\pi}\int_{\mathbb{D}^2} d^2x\;
  \left(\nabla^2u - 2\pi\sum_{i=1}^N\delta^{(2)}(z - z_i)\right)
= \frac{1}{2\pi}\int_{\mathbb{D}^2} d^2x
= \frac{1}{2} R^2,
\eeq
and so we have
\beq
R = \sqrt{2N}.
\eeq

For completeness, we evaluate the boundary term in the toy model
\eqref{eq:toymodel} for the solution \eqref{eq:flatsol_u0fixed2}
\begin{align}
  &-i\int_{D} d^2x \; \epsilon^{ab}\p_a(\bar{\phi}D_b\phi) 
  = 2\int_{D} d^2x \;\p_{\bar{z}}(e^{2u}\p_z u) \non
  &= -i\oint_{\p D} dz\;
  e^{\frac{R^2-|z|^2}{2}}\prod_{j=1}^N\frac{|z-z_j|^2}{R^2}
  \left(-\frac{\bar{z}}{4} + \sum_{i=1}^N\frac{1}{2(z-z_i)}\right).
  \label{eq:bdryterm_toymodel}
\end{align}
Notice that the conformal factor of the metric drops out and the
result does not depend on the metric (but only on the shape of the 
integration domain).
For $N=1$, the expression is fairly simple and yields
\beq
2\pi\left[\left(1 + \frac{|z_1|^2}{R^2}\right)
  \left(-\frac{R^2}{4} + \frac{1}{2}\right) -
  \frac{|z_1|^2}{2R^2}\right]
= - \frac{\pi |z_1|^2}{2},
\eeq
where we have used the Bradlow integral relation $R=\sqrt{2N}$ in the
last equality.
One could interpret this result as the boundary attracting the single
vortex and only at the center of the disc the attraction cancels out
(i.e.~it becomes isotropic). 
For general vortex positions, the contribution
\eqref{eq:bdryterm_toymodel} is in general a complicated function. 
However, setting $z_i=0$, we get the axially symmetric solution
\eqref{eq:Bradlow_radial_BC} for which the boundary contribution 
reduces to
\beq
-i\oint_{\p D} dz\;
\left(-\frac{R^2}{4z} + \frac{N}{2z}\right)
=2\pi\left(-\frac{R^2}{4} + \frac{N}{2}\right) = 0,
\eeq
where we have used the Bradlow integral relation $R=\sqrt{2N}$.
Let us emphasize that the boundary term depends on the explicit choice
of underlying theory and this is just a concrete example of the toy
model studied in Sec.~\ref{sec:toymodel}.

\subsection{Bradlow vortices on $M_0$ with nontrivial metric}

In this section we extend the Bradlow vortex solutions to nontrivial
manifolds of nonvanishing and non-constant curvature. 
Let us consider metrics of the form 
\beq
ds^2 = dt^2 - \Omega_0(|z|^2) dz d\bar{z},
\eeq
where the conformal factor, $\Omega_0$ depends only on the modulus 
(squared) of the complex coordinate $|z|^2$. 
From the Bradlow equation \eqref{eq:Bradlow} we can see that the
solution $u$ has two contributions; one is the logarithmic terms
corresponding to the vortex positions (zeros of $e^{2u}$) and the
other is the inverse Laplacian of the conformal factor $\Omega_0$. 
Formally, we can write the solutions as
\beq
u = u_0 - F(|z|^2)
+ \frac{1}{2}\sum_{i=1}^N\log|z-z_i|^2
+ g(z) + \overline{g(z)},
\label{eq:uBradlowsol}
\eeq
where the function $F$ is the solution to
\beq
\nabla^2 F = \Omega_0.
\eeq
Now imposing the boundary condition $u(z_{\p M_0})=0$, i.e.~$u$
vanishes at the boundary of $M_0$ is highly nontrivial.
If we choose $M_0$ to be a disc with conformal factor $\Omega_0$ and
radius $R$, then we can again use Conjecture \ref{conjecture1} to set
all $z_i=0$ and hence the solution satisfying the boundary condition
$u(R)=0$ reads
\beq
u = F(R^2) - F(|z|^2)
+ \frac{N}{2}\log\frac{|z|^2}{R^2}.
\label{eq:uBradlowsol_BC}
\eeq

\subsection{Solutions for a class of metrics}

Let us now consider a class of metrics of the form
\beq
ds^2 = dt^2 - \kappa^{-1} (1 \pm |z|^{2k})^\ell dz d\bar{z},
\label{eq:metricclass}
\eeq
with $k\in\mathbb{Z}_{>0}$ a positive definite integer, 
$\ell\in\mathbb{Z}$ is an integer, $\kappa\in\mathbb{R}_{>0}$ is a
real positive-definite constant. For the lower sign, the coordinate 
$z$ is defined within the unit circle: $|z|<1$ and for the upper sign
$z\in\mathbb{C}$.   
$\ell=0$ corresponds to the flat disc $\mathbb{D}^2$.
Another special case is $\ell=-2$ and $k=1$ which is the 2-sphere
$S^2$ for the upper sign and the hyperbolic plane $\mathbb{H}^2$ for
the lower sign, both with constant Gaussian curvature.
To see this, let us calculate the Gaussian curvature for this manifold
\beq
K_0 = -\frac{1}{2\Omega_0}\nabla^2\log\Omega_0
= \mp \frac{2\kappa\ell k^2 |z|^{2k-2}}{(1 \pm |z|^{2k})^{\ell+2}}.
\eeq
As promised, $\ell=-2$ renders the denominator constant and $k=1$ the
numerator constant, yielding a constant positive (negative) Gaussian
curvature of $4\kappa$ ($-4\kappa$) for the upper (lower) signs. 
If $k>1$ then the curvature vanishes at the origin but is
non-vanishing away from it.
If $k=1$ then only the denominator influences the curvature and hence
increases (decreases) the curvature with increasing radii $|z|$ for
$\ell>0$ or $\ell<-2$ (for $\ell=-1$) for both signs, while the
curvature is constant and vanishing for $\ell=0$ and just constant for
$\ell=-2$. 

The analytic solutions to the Bradlow equation in this case are thus
given by Eq.~\eqref{eq:uBradlowsol} with the $F$-function for this
class of metrics, i.e.,
\beq
F^{(\ell,k)} = \frac{|z|^2}{4\kappa}
{}_3F_2\left[k^{-1},-\ell,k^{-1};1+k^{-1},1+k^{-1};\mp |z|^{2k}\right],
\label{eq:Fsol}
\eeq
where ${}_3F_2$ is a hypergeometric function.

In some cases, we can write $F$ as a (finite) sum of fractions. 
As a good check, let us first consider the solution for $\ell=-2$ and
$k=1$ for which we know that the Gaussian curvature is constant.
In that case, we get
\beq
F^{(-2,1)} = \pm\frac{1}{4\kappa}\log(1 \pm |z|^2),
\eeq
as one would expect \cite{Manton:2016waw} (in the latter reference
$\kappa=1/4$). 
Note that this function does not have any singularities inducing more 
vortices as $|z|<1$ for the lower sign.

Another sanity check of the solution \eqref{eq:Fsol} is $\ell=0$ for
which it reduces to
\beq
F^{(0,k)} = \frac{|z|^2}{4\kappa},
\eeq
which is the solution for the flat disc $\mathbb{D}^2$, see
Eq.~\eqref{eq:flatsol} (the latter equation corresponds to
$\kappa=1$). 

Other families of solutions that can be written as fractions are, for 
$\ell=1$:
\beq
\kappa F^{(1,k)} = \frac{|z|^2}{4} \pm \frac{|z|^{2k+2}}{4(1+k)^2},
\eeq
and for $\ell=2$:
\beq
\kappa F^{(2,k)} = \frac{|z|^2}{4}
\pm \frac{|z|^{2k+2}}{2(1+k)^2}
+ \frac{|z|^{4k+2}}{4(1+2k)^2},
\eeq
and for generic $\ell\geq 1$:
\beq
F^{(\ell\geq 1,k)} =
\frac{|z|^2}{4\kappa}\sum_{p=0}^{\ell}
\left(\begin{array}{c} \ell\\ p\end{array}\right)
\frac{(\pm 1)^p|z|^{2 p k}}{(1 + p k)^2}.
\eeq
Finally, for $\ell=-1$ we can write the solution as
\beq
F^{(-1,k)} = \frac{|z|^2}{4\kappa k^2}\Phi\left[\mp |z|^{2k},2,k^{-1}\right]
= \frac{|z|^2}{2\kappa}
\sum_{p=0}^{\infty} \frac{(\pm 1)^p|z|^{2pk}}{(1 + p k)^2},
\eeq
where $\Phi$ is the Hurwitz-Lerch transcendent.
The right-most expression is only well defined for $|z|<1$.

\subsection{Flux matching}

We have found the analytic solution in closed form for the class of
metrics \eqref{eq:metricclass}; however, from a physical point of
view, we should still make sure that the number of vortices $N$
specified with positions by the delta functions in
Eq.~\eqref{eq:Bradlow} match the magnetic flux, i.e.~that
\beq
2\pi N = \int_{M_0} d^2x\; \Omega_0 = A,
\label{eq:Bradlow_bound2}
\eeq
holds, which is simply Eq.~\eqref{eq:NArel}. 
By Green's theorem the magnetic flux is giving the vortex number, $N$, 
but in order for it to match the area of the manifold, $M_0$, the
above equation needs to be imposed as well.
For the upper sign (in the solutions) with $\ell\geq 0$ or for the
lower sign with $\ell<0$ this just restricts the size of the domain
where the vortices have support. 
For the upper sign with $\ell<0$ or for the lower sign with $\ell>0$
this relation can restrict the number of vortices similarly to the
Bradlow bound, depending on the values of $\ell$ and $k$.

Let us calculate the area for the metrics \eqref{eq:metricclass} in
the case of an axially symmetric domain (for simplicity)
\beq
A^{(\ell,k)} = \frac{2\pi}{\kappa}\int_0^R dr\; r(1 \pm r^{2k})^\ell
= \frac{\pi R^2}{\kappa}
{}_2F_1\left[k^{-1},-\ell;1+k^{-1}; \mp R^{2k}\right],
\eeq
where ${}_2F_1$ is the Gaussian hypergeometric function.
As a consistency check, we can set $\ell=0$ and verify that
\beq
A^{(0,k)} = \frac{\pi R^2}{\kappa},
\eeq
as it should for a flat disc, $\mathbb{D}^2$.
We can again simplify the hypergeometric function in cases of positive 
$\ell$; in particular for $\ell=1$:
\beq
A^{(1,k)} = \frac{\pi R^2}{\kappa}\left(1 \pm \frac{R^{2k}}{1+k}\right),
\eeq
and for $\ell=2$:
\beq
A^{(2,k)} = \frac{\pi R^2}{\kappa}\left(1 \pm \frac{2R^{2k}}{1+k} +
\frac{R^{4k}}{1+2k}\right),
\eeq
and for generic $\ell\geq 1$:
\beq
A^{(\ell,k)} = \frac{\pi R^2}{\kappa}
\sum_{p=0}^{\ell}
\begin{pmatrix} \ell\\ p\end{pmatrix}
  \frac{(\pm 1)^p R^{2 p k}}{1 + p k}.
\eeq  
As mentioned above, for the upper sign with $\ell\geq 0$, this area
just fixes the radius in terms of the vortex number by
Eq.~\eqref{eq:Bradlow_bound2}, but for the lower sign with $\ell>0$ it
can limit the number of vortices possible. 

Let us consider the lower sign with the radius $R=1-\epsilon$, where
$\epsilon$ is an infinitesimal real number.
In this case, we can expand the Gaussian hypergeometric function to
get 
\beq
A^{(\ell,k)} =
-\frac{\pi^2 R^2\csc(\pi\ell)\Gamma\left(1+k^{-1}\right)}
{\kappa\Gamma(-\ell)\Gamma\left(1+k^{-1}+\ell\right)}(1+2\epsilon)
+\epsilon^\ell\left(-\frac{2^{1+\ell}k^\ell \pi r^2}{\kappa(1+\ell)}\epsilon
+\mathcal{O}(\epsilon^2)\right).
\eeq
We can see that if $\ell<-1$ then the second term diverges and thus
yields an unlimited area (as expected).
In the case of $\ell=-1$, $\csc(-\pi)$ is ill-defined (infinite) and
the area is again infinite.
We thus confirmed that for the lower sign with $\ell<0$, the area is
unlimited for $R<1$.
However, for $\ell\geq 0$, the area renders finite and as a few
examples we get for $R=1$
\begin{align}
  A^{(0,k)} &< \frac{\pi}{\kappa}, \\
  A^{(1,k)} &< \frac{\pi}{\kappa}\frac{\Gamma(1+k^{-1})}{\Gamma(2+k^{-1})}, \\
  A^{(2,k)} &< \frac{2\pi}{\kappa}\frac{\Gamma(1+k^{-1})}{\Gamma(3+k^{-1})},
\end{align}
and for general $\ell\geq 0$:
\beq
A^{(\ell\geq 0,k)} < \frac{\ell!\pi}{\kappa}
\frac{\Gamma(1+k^{-1})}{\Gamma(\ell+1+k^{-1})}.
\label{eq:Aellk}
\eeq
Since the right-hand side is a monotonically increasing function with
$k$, the hardest restriction happens when $k=1$ for which we can write
\beq
A^{(\ell\geq 0,1)} < \frac{\ell!\pi}{\kappa}\frac{\Gamma(2)}{\Gamma(2+\ell)}
= \frac{\pi}{\kappa(1+\ell)}.
\eeq
We can write this as a Bradlow bound
\beq
N < \frac{1}{2\kappa(1+\ell)},
\eeq
which requires $\kappa<1/2$ to allow for a single vortex and even
smaller for $\ell>0$.
The biggest areas we can get from Eq.~\eqref{eq:Aellk} is by sending
$k\to\infty$ for which we get
\beq
A^{(\ell\geq 0,\infty)} < \frac{\pi}{\kappa},
\eeq
yielding the Bradlow bound
\beq
N < \frac{1}{2\kappa}. 
\eeq

In case of the upper sign and $\ell<0$, the area is also finite and
limits the vortex number.
Let us consider $\ell=-2$, for which we get
\beq
A^{(-2,k)} = \left(1 - \frac{1}{k}\right)\frac{\pi}{\kappa}
  \Gamma\left(1 + \frac{1}{k}\right)
  \Gamma\left(1 - \frac{1}{k}\right)
= \frac{k-1}{k^2} \pi^2\csc\left(\frac{\pi}{k}\right).
\label{eq:Am2k}
\eeq
This area is maximal for the two limits: $k=1$ and $k\to\infty$: both
yielding
\beq
A^{(-2,1)} = A^{(-2,\infty)} = \frac{\pi}{\kappa},
\eeq
where the first is the area of the 2-sphere ($\kappa=1/4$ corresponds
to the unit 2-sphere) and this in turn gives the
Bradlow bound 
\beq
N \leq \frac{1}{2\kappa}.
\eeq
The most restricting bound is obtained for the smallest area of the
function \eqref{eq:Am2k}, which is for $k=2$:
\beq
A^{(-2,2)} = \frac{\pi^2}{4\kappa},
\eeq
and in turn the Bradlow bound
\beq
N \leq \frac{\pi}{8\kappa}.
\eeq
For $k=2$ to the limit $k\to\infty$, the area \eqref{eq:Am2k} grows
monotonically with $k$. 

We have thus shown that vortices can exist for any finite $\kappa$
when $\ell<0$ for the lower sign and for small enough $\kappa$ and
$\ell\geq 0$, again for the lower sign.
For the upper sign, there is no restriction on the vortex number when
$\ell\geq 0$, but for $\ell<0$ the Bradlow bound again limits the vortex
number; again vortices can only exist for small enough $\kappa$.

\section{Discussion}\label{sec:discussion}

In this paper we have constructed a two-parameter family of new
analytic solutions to the newly discovered Bradlow equation for a
special kind of vortices.
The derivation of the equation relies on the Bogomol'nyi trick and
thus gives a single second order PDE for the vortices; this implicitly
means that they are critically coupled \cite{Manton:2016waw}.
From the same action giving rise to said equations, the vortex scalar
field does not contribute to the energy; only the magnetic field and
the constant corresponding to the vacuum expectation value 
of the scalar field appear.
We would like to think of this as a system in which the magnetic field
dominates and in the same time contains vortices that are
energetically negligible.
If such system -- if only approximately -- can be realized
experimentally, our solutions may find use there.
Bose-Einstein condensates (BECs) with constant magnetic fields can be
realized experimentally by trapped ultracold atomic gases, for which
these magnetic fields are optically synthesized although the trapped
atoms are neutral \cite{Lin:2009}; if the magnetic fields are larger
than a critical value, vortices -- but global vortices -- are created,
where the magnetic field remains constant even in the presence of
vortices, which is in good agreement with the Bradlow equation.
Then, the question is whether the vortices contribute negligibly to
the total energy.  
Although this may not be true for BECs, we hope that 
it may be described approximately by the Bradlow equation.
It is also possible that a potential 
trapping atoms may be designed to have minima on  
a curved two-dimensional surface 
so that a curved space is realized.
Finally, it is perhaps possible that the materials 
in experiment are
genuinely curved; for this, some metric can easily be constructed (if
not already in our class of metrics) and the Bradlow vortex can
probably readily be calculated. 
Even if the flat metric is the one that finds use in any experimental
setup, then we also provide such solution, for the first time. 

In some sense the Bradlow vortex is somewhat similar to the interior
of the large-winding Bolognesi vortex -- up to a constant proportional
to the vortex potential times the area \cite{Bolognesi:2005zr}. 
The vortex condensates of these two systems, however, obey different
dynamics, of course. 

Finally, for the Taubes equation, 
a non-Abelian extension is possible which is 
most easily achieved by using the moduli matrix technique 
\cite{Eto:2006pg},
for which the so-called the master equation reduces 
to the Taubes equation for the U(1) case.
A natural question is whether there exists a non-Abelian extension 
for the case of the Bradlow equation or other types of equations 
mentioned in the introduction.

\subsection*{Acknowledgments}

S.~B.~G.~thanks the Recruitment Program of High-end Foreign
Experts for support.
The work of S.~B.~G.~was supported by the National Natural Science
Foundation of China (Grant No.~11675223).
The work of M.~N.~is supported in part by a Grant-in-Aid for
Scientific Research on Innovative Areas ``Topological Materials
Science'' (KAKENHI Grant No.~15H05855) and ``Nuclear Matter in Neutron
Stars Investigated by Experiments and Astronomical Observations''
(KAKENHI Grant No.~15H00841) from the the Ministry of Education,
Culture, Sports, Science (MEXT) of Japan. The work of M.~N.~is also
supported in part by the Japan Society for the Promotion of Science
(JSPS) Grant-in-Aid for Scientific Research (KAKENHI Grant
No.~25400268) and by the MEXT-Supported Program for the Strategic
Research Foundation at Private Universities ``Topological Science''
(Grant No.~S1511006).

\appendix

\section{Uniqueness}

Let us consider uniqueness of the vortex equations \eqref{eq:vtx}. Let
us start by assuming that two \emph{different} solutions to the
same equation exist, $u_{1,2}$ and both having exactly the same vortex
positions (same moduli) and both satisfy the same boundary condition
\eqref{eq:BC} appropriate for the specific equation (except for the
Ambj\o rn-Olesen-Manton equation, for which another boundary condition
should be specified). 
We define
\beq
\delta u \equiv u_1 - u_2,
\eeq
and subtract their two respective equations of motion which yields 
\beq
-\frac{1}{\Omega_0}\nabla^2 \delta u = C e^{2u_2}\left(e^{2\delta u} -
1\right).
\label{eq:fluc}
\eeq
This equation is independent of $C_0$ and the delta functions present
in Eq.~\eqref{eq:vtx} also canceled out.
Since the delta functions are gone, no logarithmic singularities can
be present in the solution $\delta u$ and since both $u_{1,2}$ obey
the same boundary condition, $\delta u\to 0$ asymptotically or at the
boundary of the manifold $M_0$. 

A key observation is that since $u_2\in\mathbb{R}$ is a real-valued
field, $e^{2u_2}$ is positive semi-definite and vanishes only at the
vortex centers.

Let us consider a simplified situation where we locate all vortices at
the origin of our manifold $M_0$ (in some coordinates) and to make
sure that the vortices are not destroyed, we impose $\delta u=0$ at
the vortex position. Now it is clear that since Eq.~\eqref{eq:fluc}
yields a monotonic behavior for $\delta u$; more specifically
\beq
    {\rm sign}\left[C\Omega_0\delta u\right] =
    \left\{\begin{array}{ll} +, & \delta u \ \textrm{monotonically decreasing}\\
      -, & \delta u \ \textrm{monotonically increasing}
    \end{array}\right.
\eeq
Then it is clear that no monotonically behaving function can satisfy 
$\delta u=0$ at the boundary of $M_0$ and simultaneously $\delta
u(z_i)=0,\;\forall i$.  
A more rigorous proof can be carried out along the lines of Taubes'
proof \cite{Taubes:1979tm}, which was made for the case $C_0=C=-1$.

In the case of the Bradlow equation \eqref{eq:Bradlow}, $C=0$ and
hence we have that the covariant Laplacian of the perturbation $\delta u$
on $M_0$, vanishes
\beq
\frac{1}{\Omega_0}\nabla^2 \delta u = 0.
\eeq
It is clear that no regular nontrivial solution with $\delta u=0$ at
the boundary exists. Therefore, the Bradlow vortex is unique once the
moduli and boundary conditions have been specified (the boundary
conditions completely fixes the part of the homogeneous solution).

\end{document}